\documentclass[12pt,preprint]{aastex}
\usepackage{natbib}

\shorttitle{CFW Model for Eta Car}
\shortauthors{Soker}
\slugcomment{Draft version of \today}

\begin{document}

\title{Accretion-Induced Collimated Fast Wind Model for Eta Carinae}

\author{Noam Soker\\ 
Department of Physics, Oranim, Tivon 36006, Israel;
soker@physics.technion.ac.il.}

\begin{abstract}

I propose a scenario to account for the fast polar outflow
detected to be blown by $\eta$ Carinae in 2000.
The scenario accounts also for the lack of this flow in
1998 and 1999.
The scenario is based on the binary nature of $\eta$ Carinae.
The collision of the winds blown by the two stars,
{{{{{in particular near periastron passages, }}}}}  slows down
$\sim 5\%$ of the wind blown by $\eta$ Carinae, i.e.,
the massive primary star, such that it stays bound to the system. 
I assume that most of this mass is accreted back through the equator
by $\eta$ Carinae during apastron passage, when its orbital velocity
is much lower.
The mass accretion rate of this backflowing material may become
$\sim 25 \%$ of the wind's mass loss rate {{{{{ near apastron passages. }}}}}
If the backflowing matter has enough specific angular momentum it
can form an accretion disk and may lead to the formation of a polar
collimated fast wind (CFW) on top of the stellar wind.
The gaps and uncertainties in this scenario, which should be
closed and refined in future theoretical works, as well as several
predictions which can be tested with observations in
the near future, are discussed.

\end{abstract}

{\bf Key words:} binaries: close$-$circumstellar matter$-$stars:
individual: $\eta$ Carinae$-$stars: mass loss$-$stars: winds

\section{INTRODUCTION}
\label{sec:intro}
 
Observations and models that support the presence of a close binary
companion to $\eta$ Car have rapidly accumulated in recent years
(e.g., Damineli 1996; Ishibashi et al.\ 1999; Damineli et al.\ 2000;
Corcoran et al.\ 2001a,b; Pittard \& Corcoran 2002;
{{{ Duncan \& White 2003). }}}
These papers deal mainly with the luminoisty of $\eta$ Car,
either the X-ray properties (e.g., Corcoran et al.\ 2001a,b),
or the spectroscopic event$-$the fading of high excitation lines
(e.g., Damineli et al.\ 2000;
{{{ Vieira, Gull \& Danks 2003). }}}
In a previous paper (Soker 2001a) I argue that the bipolar
shape$-$two lobes with an equatorial waist between them$-$of
$\eta$ Car and its departure from axisymmetry is best explained
by the binary model.
In particular, the kinetic energy and momentum of the $\eta$ Car
nebula is more easily explained by two oppositely flowing jets,
or a collimated fast wind (CFW), which were blown by a binary
companion during the 20-year long Great Eruption of 1850.
{{{ This is even more so for the recently claimed more massive
lobes (Smith et al.\ 2003b).
The lobes have more kinetic energy, which is not easily accounted
for in a single star model.
In the binary model, the companion simply accreted more mass
during the Great Eruption, by that blowing a stronger CFW
(Soker 2001a). }}}
Other researchers prefer to attribute the bipolar shape
to axisymmetrical mass loss geometry from a single star
(e.g., Maeder \& Desjacques 2001; Smith et al.\ 2000, 2003a,
hereafter S2003).

The presence of an eccentric binary system at the center of the
bipolar $\eta$ Car nebula is not a surprise to many people
studying planetary nebulae (PNs) and symbiotic systems (Soker 2001a).
There are many symbiotic nebulae and PNs, or proto-PNs,
with bipolar structures which are similar in many respects to the
structure of the $\eta$ Car nebula, e.g.,
the symbiotic Mira Henize 2-104 (Corradi et al.\ 2001), and the
proto-PN Red Rectangle (Waters et al.\ 1998),
with its central binary system (HD 44179) of orbital period
318 days, semimajor axis of $0.3 {\rm AU}$, and eccentricity
of $e=0.38$ (Waelkens et al. 1996; Waters et al. 1998).
{{{{ Rodriguez, Corradi, \& Mampaso (2001) claim for the presence
of binary central stars in three bipolar PNs, and 
discuss their similarity to symbiotic nebulae. }}}}  
An interesting object is Menzel 3, with its bipolar main structure
and many strings (Kastner et al.\ 2003), similar in some respects
to those of $\eta$ Car (Weis, Duschl, \& Chu 1999) and which
is thought to be either a proto PN or a symbiotic binary system
(Schmeja \& Kimeswenger 2001; Zhang \& Liu 2002).
Kastner et al. (2003) tentatively argue for the presence
of an X-ray jet in one of the two lobes of Mz 3.
{{{ The structures of these nebulae are not identical to that
of $\eta$ Car. This is in accord with the expectation from the
binary model for the formation of bipolar structures in stellar
systems (Soker 2002).
There are about a hundred qualitatively different evolutionary routes
that binary progenitors of bipolar nebulae can take,
in some of them the companion is of low mass and/or
is destructed during the process, and will be difficult to detect
(Soker 2002).
This implies that there are about a hundred qualitatively different
types of bipolar PNs and related objects. 
Within each type there are quantitative differences as well.
Therefore, rarely we find two identical bipolar structures.  }}}

S2003 study the present mass loss geometry from $\eta$ Car,
and find the wind velocity and density to vary over time
and latitude.
Among other things, S2003 find the velocity and density to
depend relatively weakly on latitude in March 1998 and February 1999.
The wind speed at those times was
$v_w \sim 400-800~{\rm km}~{\rm s}^{-1}$.
In March 1998 the mass loss rate was clearly detected from latitude
$\sim 30^\circ$ (the equator is at $0^\circ$) to $\sim 80^\circ$,
while in February 1999 the mass loss was detected in
the latitude range of $\sim 50-80^\circ$.
In March 2000 the mass loss geometry was markedly different,
with wind speed at the poles ($90^\circ$) of
$v_w \simeq 1200~{\rm km}~{\rm s}^{-1}$, and a higher
mass loss rate along the poles than at lower latitudes.
S2003 proposed a scenario to account for these variations based
on a rapidly rotating star; they attributed no significant role
to a binary companion in shaping the wind.

Motivated by the recent findings of S2003, I propose an alternative
scenario (section 2) to the single-star scenario of S2003 to
explain the mass loss geometry found by them.
Based on the similarity of the nebular structure to those
of some bipolar PNs and symbiotic nebulae, the
scenario includes ingredients which were proposed in the context of
these objects, but with significant differences.
In particular I consider the role of winds collision,
taking the view that $\eta$ Car is a binary system.
    
\section{ACCRETION OF BACKFLOWING MATERIAL}
\label{accretion}

Accretion of backflowing material, i.e., gas blown by
the star is accreted by the same star,
was proposed to account for some properties of bipolar
planetary nebulae (Soker 2001b; Sahai et al.\ 2002).
In that scenario an asymptotic giant branch (AGB), or post-AGB,
star accretes from the dense material which it blew earlier. 
For the wind material to stay bound to the star, it was
shown that the wind should be dense and flow outward
at a low speed (Soker 2001b).
It is also suggested that such a wind may be formed close to the
equatorial plane as a result of gravitational influence by a binary
companion (Soker 2001b), and that the accreted backflowing material
is likely to have enough angular momentum to form an accretion disk.
Such an accretion disk may lead to the formation of two jets, or a CFW.

I propose that the primary star in $\eta$ Car also accretes
backflowing material during part of the orbit, forming
an accretion disk and blowing its own CFW.
For $\eta$ Car I propose that the bound dense and slow flow is
formed by the following process.
The {{{ continuous, more or less spherical, }}} wind blown by the
primary {{{ (the primary wind) }}} collides with the
{{{ continuous, more or less spherical, }}}
wind blown by the companion {{{ (the secondary wind) }}}
and rapidly cools (e.g., Pittard \& Corcoran 2002).
If the cooling time is much shorter than the flow time, then the cold
dense gas may not reach an escape velocity from the binary system.
The ratio of the cooling time, $\tau_c$, to the flow time,
$\tau_f$, is given by equation (3) of Pittard \& Corcoran (2002);
here I give it in a different form,
{{{ but scale the properties of the primary and secondary winds
according to Pittard \& Corcoran (2002). }}}
The flow time is $\tau_f \simeq d_1/v_{w1}$, where
$d_1$ is the distance of the primary from the contact
discontinuity$-$where the two winds collide$-$and $v_{w1}$ is
the terminal speed of the wind blown by the primary.
For the cooling rate I approximate the cooling curve
around a shock temperature of $T \sim 3 \times 10^6$ K
(e.g., Gaetz, Edgar, \& Chevalier 1988) by
$\Lambda = 2.5 \times 10^{-23} (T/2\times 10^7~{\rm K})^{-1/2}
~{\rm erg}~{\rm cm}^3~{\rm s}^{-1}$, for $T \lesssim 2 \times 10^7$~K.
Relating the shock temperature to the speed of a fully
ionized solar composition wind gives (similar to equation 3
of Pittard \& Corcoran 2002)
\begin{equation}
\chi \equiv \frac{\tau_c}{\tau_f} \simeq 0.01 
\left( \frac{ \dot M }{10^{-4}~M_\odot~{\rm yr}^{-1}} \right)^{-1}
\left( \frac{ d_1 }{10~{\rm AU}} \right)
\left( \frac{ v_{w1}}{500~{\rm km}~{\rm s}^{-1}} \right)^5 .
\label{eq:chi1}
\end{equation}
Near periastron $d_1 < 10 {\rm AU}$, and this ratio will be smaller,
while at apastron $d_1 \sim 20 {\rm AU}$ and the ratio is
somewhat larger, but still $\chi \ll 1$.
The primary has a varying mass loss rate (S2003).
The high mass loss rate phase is attribute to periastron passage
in the binary model; the value of $\chi$ decreases there.
The conclusion is that the cooling time, in particular near periastron,
is much shorter than the flow time.
The shocked gas will rapidly cool to $ T\sim 10^4$~K, and
be compressed by the high post-shock pressure.
Away from apastron the effect is greater, as cooling time is shorter
and shock pressure is higher.

At and near the stagnation point$-$the contact discontinuity
point along the line joining the two stars$-$the wind
velocity is perpendicular to the shock front.
As we move away from the stagnation point the angle
between the primary {{{ (or secondary) }}} wind's velocity and
the shock front depends on
the exact shape of the contact discontinuity, which
mainly depends on the ratio of the momentum flux of the
two winds (Pittard \& Corcoran 2002).
Close to the stagnation point a good approximation is that
the shock front is perpendicular to the line joining the
two star.
Let $\theta$ be the angle of the velocity of a primary's wind
segment measured from the line joining the two stars.
The velocity component parallel to the shock front does not change,
and stays at $v_p = v_{w1} \sin \theta$ behind the shock.
For a typical mass of the primary {{{ ($\sim 120 M_\odot$; }}}
Hillier et al.\ 2001),
the escape velocity (from the primary) near the stagnation point is
$v_{\rm esc} \simeq 140 (d_1/10 {\rm AU})^{-1/2}~{\rm km}~{\rm s}^{-1}$.
The shocked wind's segment will stay bound if $v_p < v_{\rm esc}$,
or 
\begin{equation}
\sin \theta < \sin \theta_b \sim
0.3  \left( \frac{d_1 }{10 {\rm AU}} \right)^{-1/2}
\left( \frac{ v_{w1}}{500~{\rm km}~{\rm s}^{-1}} \right)^{-1} .
\label{eq:theta1}
\end{equation}
This is a crude estimate, since different post-shock wind's segments
interact among themselves, and the situation is more complicated.
The fraction of post-shock material that stays bound to
the system is given by
\begin{equation}
f= \frac{1}{2} (1-\cos \theta_b) = 0.02 \qquad {\rm for} \qquad
\sin \theta_b =0.3.
\label{eq:f1}
\end{equation}
Using the binary parameters given by Corcoran et al.\ (2001a),
I find at periastron $d_1 \lesssim {2.5 \rm AU}$, hence 
$\sin \theta_b \simeq 0.5$ and $f \simeq 0.07$.
The mass loss rate increases at periastron passage.
A crude estimate gives that the fraction of the mass lost by the
primary over one orbital period and that stays bound to the system
due to its collision with the companion's wind is
$f \simeq 0.03-0.05$.
Winds collision is not the sole effect {{{ for producing bound material
near the equatorial plane.}}}
In the Red Rectangle, where no strong wind collision is expected,
there is a relatively dense bound material near the equatorial
plane outside and close to the binary system
(e.g., Jura \& Kahane 1999).
This shows that gravitational interaction may also
form a concentration of slowly moving gas and dust in the equatorial
region of a close binary system.
The exact mechanism for the formation of this ``disk'' is not clear.

  The stagnation point, where the bound gas is slowed down, is closer
to the secondary star (Pittard \& Corcoran 2002), hence it is between
the center of mass and the secondary star in $\eta$ Car.
The cold bound material is more likely to be accreted during and
near apastron, when the primary star is farther away from the
center of mass and its orbital speed is much slower.
{{{{ Note that the inflow itself starts earlier, as the system moves
away from periastron. 
The backflow time of the accreted material from a distance of
$\sim 15-20 {\rm AU}$ to the primary star is $\sim 1.5$~yr which is a
non-negligible fraction of the orbital period.
This means that if the CFW is observed near apastron, and
most of the backflowing mass originated from the massive wind
blown near periastron passage, then the backflowing material starts its
inflow $\sim 1$~yr after periastron passage. }}}}
{{{ Quantifying this process requires hydrodynamical simulations,
which are beyond the scope of the present paper.
I note the enhanced mass loss rate during the periastron passage of 1998
(Duncan \& White 2003), which also favors accretion during apastron
passage. }}}
Therefore, the accretion phase is shorter than the orbital period.
According to the present scenario this explains the finding by S2003
that only during part of the orbit is there a fast polar flow.
As illustrative values I take a fraction of $f \sim 0.05$ of the
total mass lost during an orbit to stay bound and be accreted,
I take the accretion phase to last a fraction $\eta \sim 0.3-0.5$ of the
orbital period, and I take a lower than average
mass loss rate during and near apastron.
Let $\beta \sim 0.5$ be the ratio of apastron mass loss rate to the
average over an orbit mass loss rate.
These values imply that during the accretion phase near apastron,
the ratio of accretion to mass loss rate is
\begin{equation}
\frac{ \dot M_{\rm acc}}{\dot M_{\rm ap}}
= \frac{f}{\eta \beta} \sim 0.25.
\label{eq:acc1}
\end{equation}
The value in the last equality is for $f \sim 0.05$, $\eta \sim 0.4$,
and $\beta \sim 0.5$.

{{{{ As the dense post-shock material flows back to the primary star,
three relevant forces are acting on it: gravity, wind's ram pressure,
and radiation force.
For the parameters of the present system, wind's ram pressure is somewhat
larger than radiation pressure (eq. 6 of Soker 2001b).
It is therefore adequate to consider the force due to wind's ram
pressure and gravity.
I consider a post-shock spherical blob of mass $m_b$, density $\rho_b$,
and radius $r_b$, located at a distance $r$ from the primary of mass $M_1$.
I also take the blob's density to be $B$ times the wind's density
$\rho_b = B \rho_{w1}$. 
The magnitude of the gravitational force is given by 
\begin{equation}
f_g = \frac{G M_1 m_b}{r^2} =
\frac{G M_1}{r^2} \frac {4 \pi}{3} r_b^3 B \rho_{w1} , 
\end{equation}
where in the second equality I substituted for the blob mass. 
The force due to the ram pressure of the wind is given by
\begin{equation}
f_{w1} = \rho_{w1} v^2_{w1} \pi r_b^2.
\end{equation}
The ratio of these forces is
\begin{equation}
F \equiv \frac {f_g}{f_{w1}} = \frac{4}{3}
\frac{G M_1}{r^2 v^2_{w1}}  r_b B \simeq
\frac{R_1}{r} \frac {r_b}{r} B ,
\end{equation}
where in the second equality I took the vind speed to be of the
order of the Keplerian speed on the surface of the primary star,
of radius $R_1$.
For the present system, the first factor on the far right hand side of the
last equation is $R_1/r \gtrsim 0.02$, where the minimum value is obtained
for the distance of the stagnation point at apastron, $d_1 \sim 20 {\rm AU}$. 
For pressure equilibrium between the ram pressure of the wind and
a cool blob at a temperature of $T \sim 10^4 K$, the compression factor
is $B \gtrsim 1000$.
Because the post-shock blobs were formed when mass loss rate was higher
near periastron passage, this factor can be even larger $B \sim 2000$.
The requirement for the blobs to be accreted, therefore, is that
its size be $r_b \gtrsim 0.025 r$.
This condition is likely to be met, as the region near the stagnation
point from which mass is accreted is much larger (eq. 2).
Smaller blobs may not be accreted.
For conditions a little less favorable, such a backflow
will not occur.
In the present paper I claim that in $\eta$ Car conditions are such that
backflow do occur. It does not mean that these conditions will hold
in the future, when wind properties may change.
}}}}

The accretion occurs very close to the equatorial plane, as the
dense cold gas is formed there and gravity forces it to flow there.
{{{{{ In young stellar objects it is well established that accretion
and outflow occur simultanouosly. Here, I show that the backflowing
blobs cover a small fraction of the hemisphere, untill they form an
accretion disk close to the star, and hence I expect outflow and
inflow to occur simulatanouosly near apastron passages;
near periastron passages only outflow occurs.
The average inflow rate through a radius $r$ is given by
\begin{equation}
\dot M_{\rm acc} = \Omega_{\rm in} v_{\rm in} \rho_b 4 \pi r^2,
\end{equation}
where $\Omega_{\rm in}$ is the fraction of solid angle covered by the infllowing
gas (blobs), and $v_{\rm in}(r)$ is the infflow speed at radius $r$.
The mass outflow rate near apastron passage is given by
\begin{equation}
\dot M_{\rm ap} = v_{w1} \rho_{w1} 4 \pi r^2.
\end{equation}
Dividing equation (8) by equation (9), and rearranging, gives for
the covering solid angle fraction of the inflowing gas
\begin{equation}
\Omega_{\rm in} =
\frac{\dot M_{\rm acc}}{\dot M_{\rm ap}}
\frac{ v_{w1}}{v_{\rm in}}
\frac{\rho_{w1}}{\rho_b}.
\end{equation}
From equation (4) and the discussion bellow,
$\dot M_{\rm acc} \simeq \dot M_{\rm ap}$.
From the discussion above for the compression ratio $B$,
$\rho_{w1}/\rho_b = 1/B \sim 10^{-3}$.
The inflow gas is accelerated inward, hence its speed is
bellow that of the outflowing gas, but not by much because the
stagnation point to primary radius is not a huge ratio.
Taking $v_{w1}/{v_{\rm in}} \lesssim 10$, I find
$\Omega_{\rm in} \sim 0.01$.
Therefore, the dense inflowing gas covers a small fraction
of the outflowing gas.
The accretion disk itself and the wind blown by it may disrupt
segments of the wind while the coolimated outflow is active.
In anycase, most of the backflowing gas results from wind blown near
the apastron passage.
}}}}}

{{{ In a recent paper Duncan \& White (2003) claim that
the mass loss rate in the equatorial plane was enhanced during the
periastron passage of 1998. This enhanced mass loss not only
concentrates mass to the equatorial plane, but also implies
a lower value of $\beta$ used in equation (4), enhancing the
effect proposed here. }}}
Because of the binary nature of the system it is expected
that the accreted mass has high specific angular momentum, and
an accretion disk is likely to be formed.
The accretion flow stops the wind only very close to the equatorial
plane. It is clear that the wind dominates over accretion
for the ratio found in equation (4), and {{{ a very }}}
well collimated very dense jets are not expected even if the 
accreted backflowing material forms an accretion disk.
{{{ This is so because any wind blown by the accretion disk
will interact close to the star with the wind blown from the
surface of the star. This interaction is likely to degrade
any collimation. In addition, the primary wind is not collimated,
hence reducing the density contrast between the polar CFW and
the wind in other directions. }}} 
However, it is quite possible that the accretion disk will lead
to the formation of a polar flow, added to the wind, resulting in
a faster and higher density polar flow.
{{{ Jets (or CFW)  launched by accretion disks move at about
the escape velocity from the accreting object {{{{ (Livio 2000). }}}} 
Winds blown by giant stars may move at a speed somewhat 
bellow the escape speed. If this is the case with the primary
(more or less spherical) wind, then this explain the }}} 
sharp rise in outflow speed toward the poles found by S2003 
in March 2000. 

{{{ I stress the significant differences between the presently proposed
scenario for the present flow in $\eta$ Car,
and the model proposed in Soker (2001a) to account for the
the formation of the main bipolar structure (the Homunculus) of $\eta$ Car
during the Eruptions of the 19th century.
According to Soker (2001a; see discussion following eq. 2 in that paper),
during the Great Eruption of 1850 (and possibly during the
Lesser Eruption of 1890) the mass loss rate from the massive
primary star was huge and the wind's speed lower, such that the
stagnation point of the colliding winds was {\it inside} the accretion
radius of the secondary star.
Therefore, the shocked winds' material was accreted
directly and continuously by the secondary star.
An accretion disk was formed during the entire Great Eruption period
around the secondary star, leading to the ejection of a CFW by the
secondary (lower mass) star.
According to that scenario (Soker 2001a), two winds shaped the
main bipolar structure during the Great Eruption:
A dense relatively slow wind blown by the primary star, and
a CFW blown by the {\it secondary} star.
Presently, the primary's wind mass loss rate is much lower than that
during the Great Eruption, and the stagnation point is outside the
accretion radius of the secondary star. Hence, the shocked
winds' material is not accreted by the secondary star.
As discussed here, it is more likely to be accreted later in the
orbit by the primary star.
Presently, three types of winds can be identified:
The continoues wind blown by the primary, the continues wind blown by
the secondary, and an intermittent CFW blown by the accretion disk
around the primary star during part of the orbital period. }}}

\section{DISCUSSION AND SUMMARY}
\label{summary}

I proposed a scenario to account for the $\eta$ Car wind
geometry and its time variation as found recently by S2003.
The scenario is based on the binary nature of $\eta$ Car,
and the collision between the winds blown by the two components.
The post-shocked wind blown by the primary$-$the more massive
star$-$ has a cooling time near the stagnation point
much shorter than the flow time (eq. 1).
The material which is shocked near the stagnation point loses
most of its energy via radiation, and stays bound to the
binary system, as well as to the primary star (eq. 2).
A fraction of $\sim 3-5 \%$ of the mass blown during the orbit
may stay bound to the primary (eq. 3), assuming enhanced mass loss
rate during periastron passage {{{ (Duncan \& White 2003), }}}
as a result of tidal interaction.
If most of this mass is accreted back by the primary near
its orbital apastron passage, when it is farther away from the center of
mass and moves slowly, the accretion rate can be a non-negligible
fraction of the mass loss rate (eq. 4).
This mass is accreted near the equator, and if it has enough
angular momentum it can form an accretion disk and lead
to the formation of a collimated fast wind (CFW). 
{{{ The total mass loss rate into the CFW can be $\sim 10 \%$ of
the accretion rate. Taking into account the enhanced mass loss
rate during periastron passage (Duncan \& White 2003),
i.e., $\beta \sim 0.1-0.2$ in equation (4), 
I estiamte that the CFW can increase the mass loss rate
along the polar direction, within $\sim 25 ^\circ$ from
the polar direction, by a factor of $\sim 2$. }}}
I suggest this CFW as an explanation to the wind geometry found
by S2003 in March 2000; in the proposed scenario this epoch
corresponds to the apastron phase of the orbit.
No such polar flow is observed a year and two years earlier;
according to the proposed scenario the primary's orbital speed
is too fast to accrete at a high rate.
{{{ It is important to notice the differences between the propsed
flow structure refering to the present conditions in $\eta$ Car,
and the flow structure that occured during the Eruptions of the
19th century, as proposed in Soker (2001a). These differences
are discussed in the last paragraph of section 2. }}}

The main gaps and uncertainties in the presently proposed scenario are:
(1) Is some of the wind blown by the primary indeed
accreted back by it via an equatorial backflow during some fraction
of the orbit, as suggested here?
(2) If it is, what is the mass accretion rate, which was estimated
here in equation (4)?
(3) What is the specific angular momentum of the accreted mass?
(4) If the mass and specific angular momentum of the accreted
mass are high enough to ensure the formation of a small accretion
disk, can this disk cause the polar dense and fast
flow observed by S2003?
This is a crucial ingredient in the proposed scenario.
The single star model proposed by S2003 is not free of some
open questions as well.
(1) The model requires the star to rotate fast (S2003).
Can a single evolved star rotate fast enough?
In general such stars seems to slow down
(Maeder \& Meynet 2000),
below the rotation speed required for a significant polar flow.
(2) Can such a star blow a dense fast polar flow?
In many models (see S2003), as well as in the solar wind (e.g.,
Habbal \& Woo 2001), the density anti-correlates with the wind speed.
Even S2003 note in their figure 7 that the dependence of wind speed
on latitude is not shallow as expected in rotating stars.
It resembles more a polar jet with a half opening angle
(measured from the symmetry axis) of $\sim 20-30^\circ$. 
(3) It is not clear if the time variation in the wind geometry
proposed by S2003 can indeed work; it incorporates a response of the
rotating stellar envelope to a mass loss, a process which has not
been studied in detail.

Concerning the many open questions on the nature
of $\eta$ Car central star(s) and the formation process of
its nebula, the community should consider different models,
even when they are incomplete.
The presently proposed scenario is fundamentally different from
that proposed by S2003 in that it attributes the dense and fast
polar outflow to binary interaction.
This is along the lines of my previous paper (Soker 2001a),
where the bipolar structure of the $\eta$ Car nebula was
attribute to binary interaction, where the secondary accreted
and blew a CFW.
One of the advantages of binary models is that they can
serve as a framework for a unified paradigm for the formation
of bipolar structures around many stars:
$\eta$ Car, symbiotic systems (which are known to contain binary
central systems), bipolar PNs (some of which are known to
contain binary systems, and the others are suspected to harbor
binary systems), and other systems, e.g., SN 1987A.

Some predictions of the proposed scenario can be
tested in the near future with more detailed observations.
(1) A small departure from axisymmetry is expected in the inner region.
This is because of the different in the interaction strength and type
near periastron and apastron (see Soker 2001a for large scale
departure in the outer region).
However, because the wind speed is much higher than the orbital
speed, the departure from axisymmetry is small (Soker 2001a).
(2) The dense bound material may be detected via weak emission or
absorption as slowly moving material in the inner region,
possibly even via emission from dust.
(3) The on-off phases of the CFW are periodic,
in contrast to the single star model, where the on-off phases
may be semi-periodic. 
(4) It is plausible, but not necessary, that the secondary more
compact companion will accrete part of the bound mass. In that case
very high speed $\sim 2000-3000~{\rm km}~{\rm s}^{-1}$ polar wind
may be detected.

\bigskip

{\bf ACKNOWLEDGEMENTS}

This research was supported in part by a grant from the
Israel Science Foundation.

\end{document}